\documentclass[11pt,a4paper]{article}
\usepackage{amsmath}
\usepackage{color}
\usepackage{cite}
\usepackage{graphicx}
\usepackage{amssymb}
\usepackage{float}
\usepackage{dsfont}

\newcommand{\xf}{\ensuremath{x_{5}}}
\newcommand{\xs}{\ensuremath{x_{6}}}
\newcommand{\Rf}{\ensuremath{R_{5}}}
\newcommand{\Rs}{\ensuremath{R_{6}}}
\newcommand{\phimn}{\ensuremath{\varphi^{(m,n)}}}

\newcommand{\Sf}{\ensuremath{\Sigma_{5}}}
\newcommand{\Ss}{\ensuremath{\Sigma_{6}}}
\newcommand{\irho}{\ensuremath{I_\rho}}
\newcommand{\mgut}{\ensuremath{M_{GUT}}}
\newcommand{\iag}{\ensuremath{\alpha_{GUT}^{-1}}}

\def\beq{\begin{equation}}
\def\eeq{\end{equation}}
\def\beqn{\begin{eqnarray}}
\def\eeqn{\end{eqnarray}}

\begin{document}
\begin{flushright}
OHSTPY-HEP-T-12-001 \\
\end{flushright}
\vskip 1.5cm
 \begin{center}
{\huge SU(6) GUT Breaking on a Projective Plane \\[2ex]}
\vspace*{5mm} \vspace*{1cm}
\end{center}
\vspace*{5mm} \noindent
\vskip 0.5cm
\centerline{\bf Archana Anandakrishnan, Stuart Raby}
\vskip 0.8cm
\centerline{
\em Department of Physics, The Ohio State University,}
\centerline{\em 191 W.~Woodruff Ave, Columbus, OH 43210, USA}
\vskip1.5cm

\centerline{\bf Abstract}
\vskip .3cm
We consider a 6-dimensional supersymmetric SU(6) gauge theory and compactify two
extra-dimensions on a multiply-connected manifold with non-trivial topology.
The SU(6) is broken down to the Standard Model gauge groups in two steps by an
orbifold projection (or Wilson line), followed by a Wilson line.  The Higgs
doublets of the
low energy electroweak theory come from a chiral adjoint of SU(6).  We thus have
gauge-Higgs unification.  The three families of the Standard Model can either be
located in
the 6D bulk or at 4D N=1 supersymmetric fixed points.

We calculate the Kaluza-Klein
spectrum of states arising as a result of the orbifolding. We also
calculate the threshold corrections to the coupling constants due to this tower
of states
at the lowest compactification scale. We study the regions of parameter
space of this model where the threshold corrections are consistent with low
energy physics.
We find that the couplings receive only logarithmic corrections at all scales.
This feature can be attributed to the large N=2 6D SUSY of the underlying model.
\vskip .3cm

\newpage

\section{Introduction}
A supersymmetric grand unified description\cite{Dimopoulos:1981yj,
Dimopoulos:1981zb, Ibanez:1981yh, Sakai:1981gr, Einhorn:1981sx, Marciano:1981un}
of the fundamental forces of nature has been the holy grail of particle physics
for many years now. Such a unified description would bring some order into the
chaotic world of particle representations. In addition, the many different
parameters of the Standard Model can be tied down using a grand unified
symmetry. Unfortunately, most such unified descriptions in 4-dimensions are
haunted by many issues. Two notable problems with 4-dimensional
supersymmetric grand unified theories (SUSY GUTs) include the Higgs
doublet-triplet splitting problem and the complicated potentials required to
break the grand unified symmetry down to the Standard Model gauge groups. Apart
from these theoretical hindrances, the major setback to 4D SUSY GUTs came with
the experimental non-observation of proton decay at the predicted life-times of
the models\cite{Dermisek:2000hr,Murayama:2001ur}. SuperK places the current
lower bound on the proton lifetime ($p \rightarrow e^+ \pi^0$) to be 1 $\times \
10^{34}$ years\cite{ichep10}.
Also, SUSY GUTs, with the standard CMSSM scenario of SUSY breaking, require
GUT-scale threshold corrections of about $-$3\% in order to fit the low energy
value of the strong coupling.

An elegant and definitive solution to the above stated theoretical issues was
proposed in models of orbifold GUTs. Grand unified theories constructed in
higher dimensional spaces could be reduced to 4-dimensional GUTs by
compactifying the extra-dimensions on specific
manifolds\cite{Dienes:1998vh,Dienes:1998vg}. By doing so, it was found that the
Higgs doublet-triplet problem could be solved in a simple manner by choosing the
correct parities along the strong and the weak
directions\cite{Kawamura:2000ev,Hall:2001pg}. Many orbifold GUTs have been
constructed since then\cite{Contino:2001si, Asaka:2001eh, Hebecker:2001wq, Hall:2001zb,Dermisek:2001hp,Hall:2001xb,Kim:2002im}, with
interesting
phenomenology and realistic supersymmetric spectrum. The Kaluza-Klein tower of
states
that arise in these extra-dimensional GUTs can also account for the GUT scale
threshold
corrections\cite{Hall:2001pg,Hall:2001xb,Kim:2002im,Dundee:2008ts, Anandakrishnan:2011zn}.
It must be pointed out that orbifold GUT model-building in field theory
constructions
mirrored the earlier work in heterotic string theory
constructions\cite{Dixon:1985jw, Dixon:1986jc, Breit:1985ud, Ibanez:1987sn}.  In
recent years, many different features of orbifold compactifications have been
studied both from a phenomenological bottom-up approach, as well as top-down
from string theory.  Within the context of string theory, gauge coupling
unification occurs at the string scale.  This may occur with or without an
intermediate GUT.  However, such theories have the problem that the string scale
is typically about 20 times larger than the 4D GUT scale
\cite{Kaplunovsky:1987rp,Kaplunovsky:1993rd,Dixon:1990pc}.  One might hope that
gauge coupling unification can be reconciled with string unification by lowering
the string unification scale. It has been argued that non-local breaking of the
GUT symmetry via Wilson lines on an anisotropic orbifold can solve the problem
of string unification \cite{Ross:2004mi,Hebecker:2004ce,Trapletti:2006xv}.
In this paper we provide a self-consistent test of this hypothesis on a
particular 6D orbifold. It is possible that this orbifold GUT is an effective
low energy theory of some string compactification, but we have not come across
any compactification that would lead to an orbifold with the topology discussed
here.

In this work, we present a 6D model with SU(6) gauge symmetry and N=2
supersymmetry.  In terms of 4D language, such a 6D theory with N=2 SUSY contains
one vector adjoint and three chiral adjoints. The model has gauge-Higgs
unification with the Higgs doublets coming from one of the chiral adjoints. The
group SU(6) is broken to SU(5) $\times$ U(1)$_X$ via orbifold boundary
conditions.  Then SU(5) is broken to the Standard Model gauge group and, at the same time, Higgs
doublet-triplet splitting is accomplished by a non-local Wilson line.  The two
extra-dimensions are compactified on an orbifold that can be characterized as a
sphere with a cross-cap, as described in
\cite{Hebecker:2003we,Hebecker:2004ce,Trapletti:2006xv}. Quarks and
leptons, and their respective Yukawa couplings to the Higgs, are localized at
the orbifold fixed points which only retain an N=1 SUSY in 4D with SU(5)
$\times$
U(1)$_X$ gauge invariance (see for example Refs. \cite{Hall:2001pg, Hall:2001zb}
where this phenomenon has been discussed).

The details of the orbifold and the symmetry breaking are
discussed in Section \ref{orbifold}. We break the SU(6) $\rightarrow$ SU(5)
$\times$ U(1)$_X$
using one of the orbifold projections, locally at the fixed points. We then
break the SU(5) $\rightarrow$
SU(3) $\times$ SU(2) $\times$ U(1)$_Y$ using a Wilson line along the fifth and
sixth directions. In Section
\ref{unification}, we analyze gauge coupling unification in the SU(6) GUT model
constructed on such an orbifold and calculate the GUT-scale threshold
corrections in this scenario.
We find that unlike in most popular models of orbifold GUTs, the couplings do
not receive any power law corrections
above the compactification scale due to the effective N=4 SUSY in 4D. We analyze
the GUT-scale threshold corrections  to determine if they are at the required
level to match low energy physics. We point out that an example of an orbifold
GUT from a 6D SU(6) was considered in\cite{Hall:2001zb} with the the similar
feature of gauge-Higgs unification.  The extra-dimensions were compactified on
$T^2/(Z_2 \times Z'_2)$ and the authors obtain realistic phenomenology with
local GUT breaking. The 6D GUT theory also had an N=2 supersymmetry. As a
consequence, the coefficient of the effective 6D quadratic power law dependence
of the gauge couplings vanished, but due to the existence of fixed lines the
effective 5D linear dependence remained.

\section{GUT breaking}
\label{orbifold}
\subsection{Real Projective Plane}
\begin{figure}[ht!]
\centering
\includegraphics[width=14cm]{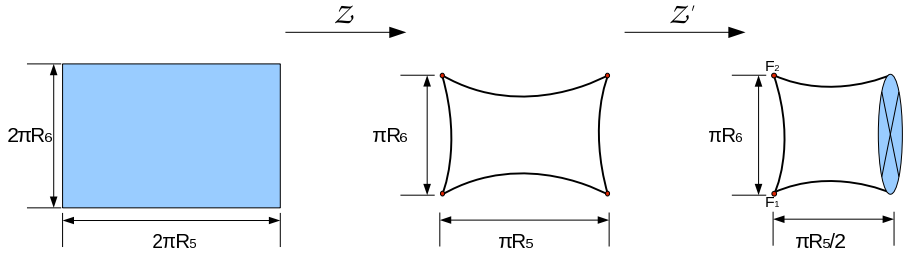}
\vspace{-35pt}
\caption{The figure shows the manifold at each step of the compactification.
After the first step of orbifolding, the space looks like a pillow with four
fixed points denoted by red dots in the center figure. After the second step of
orbifolding as described in \cite{Hebecker:2003we}, this space is equivalent to
a real projective plane.}
\label{hebecker}
\end{figure}

An N=2 supersymmetric SU(6) gauge theory in 6 dimensions is compactified on an
orbifold, shown in Fig \ref{hebecker}, as described in
Hebecker\cite{Hebecker:2003we}. The extra dimensions are compactified on a torus
$T^2$ parametrized by (\xf, \xs). The two dimensions are also identified to have
the periodicity, $x_{(5,6)} = x_{(5,6)} + 2 \pi R_{(5,6)}$, where \Rf\ and \Rs\
are the radius of the torus along the two directions. Two discrete symmetries,
the rotation ${\cal Z}$ and a freely acting roto-translation ${\cal Z}'$, as
defined in Eq.(\ref{P}, \ref{Pp}) are modded out. Once the first symmetry is
modded out, the topology of the compact space is that of a 2-sphere with
curvature concentrated at the four conical singularities. The space resembles a
pillow with fundamental group $\pi_1 = \emptyset$. Once the second parity is
modded out, the resulting compact space is equivalent to a projective plane,
$RP^2$. It is non-orientable with no boundaries, the curvature is concentrated
at the two fixed points denoted by $F_1$ and $F_2$ and $\pi_1 = {\mathbb Z}_2$.
The non-orientability of the space can be ascribed to the cross-cap where
opposite points on the circle are identified.
\begin{align}
\label{P}
{\cal Z} &\qquad& \xf \rightarrow -\xf, &\quad& \xs \rightarrow  -\xs \\
\label{Pp}
{\cal Z}'&\qquad& \xf \rightarrow -\xf + \pi \Rf, &\quad& \xs \rightarrow \xs +
\pi \Rs.
\end{align}
We choose to write the particle content of the theory in terms of the 4D
language. There is one vector superfield, V and three chiral superfields, \Sf,
\Ss, and $\Phi$. Using the notation in \cite{Hall:2001zb}, the bulk action in
the
Wess-Zumino gauge is given by:
\begin{small}
\beqn
S &=& \int d^6 x \Biggl\{ {\rm Tr} \Biggl[ \int d^2\theta \Biggl( \frac{1}{4 k
g^2} {\cal W}^\alpha {\cal W}_\alpha \nonumber \\ && \qquad + \frac{1}{k g^2}
\Bigl( \Phi \partial_5 \Sigma_6 - \Phi \partial_6 \Sigma_5 - \frac{1}{\sqrt{2}}
\Phi [\Sigma_5, \Sigma_6] \Bigr) \Biggr) + {\rm h.c.} \Biggr] \nonumber \\
 && \qquad +\int d^4\theta \frac{1}{k g^2} {\rm Tr} \Biggl[(\sqrt{2} \partial_5
+ \Sigma_5^\dagger) e^{-V}
  (-\sqrt{2} \partial_5 + \Sigma_5) e^{V} +  \nonumber \\ && \qquad (\sqrt{2}
\partial_6 + \Sigma_6^\dagger) e^{-V}
  (-\sqrt{2} \partial_6 + \Sigma_6) e^{V} \nonumber \\ && \qquad + \Phi^\dagger
e^{-V} \Phi e^{V} + \partial_5 e^{-V} \partial_5 e^{V} + \partial_6 e^{-V}
\partial_6 e^{V} \Biggr] \Biggr\}
\label{eq:5daction}
\eeqn
\end{small}
\subsection{SU(6) $\rightarrow$ SU(5) $\times$ U(1)$_X$}
The 6D N=2 supersymmetric theory that we start with has an effective N=4 SUSY in
4 dimensions.
The action of the above discussed parities can be used to break the gauge group
SU(6) down to
SU(5) $\times$ U(1)$_X$, and at the same time break N = 4 SUSY to N = 1 SUSY (in
4D) \cite{Mirabelli:1997aj}.
We can break SU(6) to SU(5) $\times$ U(1)$_X$ by requiring the fields to
transform as illustrated
below, under the two parities.\\[10pt]

\begin{small}

Under the parity, ${\cal Z}$:
 \beqn
 V(-\xf, -\xs) & =& P V(\xf, \xs) P^{-1}, \nonumber \\
\Sf(-\xf, -\xs) & =&  -P \Sf(\xf, \xs) P^{-1}, \nonumber \\
\Ss(-\xf, -\xs) &=& -P \Ss(\xf, \xs) P^{-1},\nonumber  \\
\Phi(-\xf, -\xs) &=& P \Phi(\xf, \xs) P^{-1},
\label{Ptransform}
\eeqn

Under the parity, ${\cal Z'}$:
\beqn
V(-\xf + \pi \Rf, \xs + \pi \Rs) &=&  V(\xf, \xs),  \nonumber \\
\Sf(-\xf + \pi \Rf, \xs + \pi \Rs) &=& - \Sf(\xf, \xs),  \nonumber \\
\Ss(-\xf + \pi \Rf, \xs + \pi \Rs) &=&  \Ss(\xf, \xs),  \nonumber \\
\Phi(-\xf + \pi \Rf, \xs + \pi \Rs) &=& - \Phi(\xf, \xs).
\label{Pptransform}
\eeqn
\end{small}

where P = diag$(i,i,i,i,i,-i)$, breaks the SU(6) $\rightarrow$ SU(5) $\times$
U(1)$_X$. The projection
${\cal Z}$ has four fixed points(as shown in Fig \ref{hebecker}) and hence the
SU(6) symmetry is broken
down to SU(5) $\times$ U(1)$_X$ only at those fixed points. The symmetry
breaking in this case is
said to be localized. On the other hand, the second parity is freely acting
(without any fixed points).
Therefore, breaking the gauge symmetry using the second orbifold projection
would have led to non-local
breaking of the SU(6). It can be shown that the gauge symmetry breaking by this
orbifold action
can be rewritten as symmetry breaking by a Wilson line. However, as we shall see
in the next section,
we require an additional Wilson line to further break the SU(5) down to SU(3)
$\times$ SU(2) $\times$ U(1)$_Y$.
The conditions on the Wilson lines on this orbifold (to be discussed in the next
section) do not allow
for a minimal execution of the gauge symmetry breaking from SU(6) $\rightarrow$
SU(5) $\times$ U(1)$_X$ $\rightarrow$
SU(3) $\times$ SU(2) $\times$ U(1)$_Y$ $\times$ U(1)$_X$ in a completely
non-local way.
Hence we choose to break the SU(6) $\rightarrow$ SU(5) $\times$ U(1)$_X$ locally
and the SU(5) non-locally using a Wilson line.

Under the combined operation (${\cal Z}, {\cal Z}'$) the components of the
fields transform as follows:\\[15pt]
\begin{footnotesize}
\begin{minipage}[l]{0.5\linewidth}
 \begin{align}
V = \left( \begin{array}{c|c|c}
(++) (++)  (++) & (++)  (++) & (-+) \\
(++) (++)  (++) & (++)  (++) & (-+) \\
(++)  (++)  (++) & (++)  (++) & (-+) \\ \hline
(++)  (++)  (++) & (++)  (++) & (-+) \\
(++)  (++) (++) & (++)  (++) & (-+) \\ \hline
(-+)  (-+) (-+) & (-+)  (-+) & (++) \\
 \end{array} \right) \nonumber \\
\Sigma_5 = \left( \begin{array}{c|c|c}
(--)  (--)  (--) & (--)  (--) & (+-) \\
(--)  (--)  (--) & (--)  (--) & (+-) \\
(--)  (--)  (--) & (--)  (--) & (+-) \\ \hline
(--)  (--)  (--) & (--)  (--) & (+-) \\
(--)  (--)  (--) & (--)  (--) & (+-) \\ \hline
(+-)  (+-)  (+-) & (+-)  (+-) & (--) \\
 \end{array} \right) \nonumber
\end{align}
\end{minipage}
\begin{minipage}[r]{0.5\linewidth}
 \begin{align}
\Sigma_6 = \left( \begin{array}{c|c|c}
(-+)  (-+)  (-+) & (-+)  (-+) & (++) \\
(-+)  (-+)  (-+) & (-+)  (-+) & (++) \\
(-+)  (-+)  (-+) & (-+)  (-+) & (++) \\ \hline
(-+)  (-+)  (-+) & (-+)  (-+) & (++) \\
(-+)  (-+)  (-+) & (-+)  (-+) & (++) \\ \hline
(++)  (++)  (++) & (++)  (++) & (-+) \\
 \end{array} \right) \nonumber \\
\Phi = \left( \begin{array}{c|c|c}
(+-)  (+-)  (+-) & (+-)  (+-) & (--) \\
(+-)  (+-)  (+-) & (+-)  (+-) & (--) \\
(+-)  (+-)  (+-) & (+-)  (+-) & (--) \\ \hline
(+-)  (+-)  (+-) & (+-)  (+-) & (--) \\
(+-)  (+-)  (+-) & (+-)  (+-) & (--) \\ \hline
(--)  (--)  (--) & (--)  (--) & (+-) \\
 \end{array} \right) \nonumber \\
\end{align}
\end{minipage}
\end{footnotesize}
\vspace{10pt}

The parity operations (${\cal Z}, {\cal Z}'$) performed on the coordinate space
are symmetries of the Lagrangian, hence the fields in the Lagrangian must be
eigenstates of the parity operations. A general field $\varphi = \{V,
\Sf, \Ss, \Phi\}$ by definition of the manifold, are periodic functions
of \xf\ and \xs.
\beqn
\varphi (x, \xf + 2 \pi \Rf,\xs) = \varphi (x, \xf,\xs) \nonumber \\
\varphi (x, \xf,\xs + 2 \pi \Rs) = \varphi (x, \xf,\xs)
\label{periodicity}
\eeqn

This allows us to expand them as:
\beq
 \varphi (x, \xf, \xs) = \frac{1}{\sqrt{2 \pi \Rf \Rs}} \sum_{m, n = -\infty}^{+
\infty} \phimn \text{exp} \left[ i \left( \frac{m\xf}{\Rf} + \frac{n\xs}{\Rs}
\right) \right] \\
\label{expansion}
\eeq

\begin{figure}[ht]
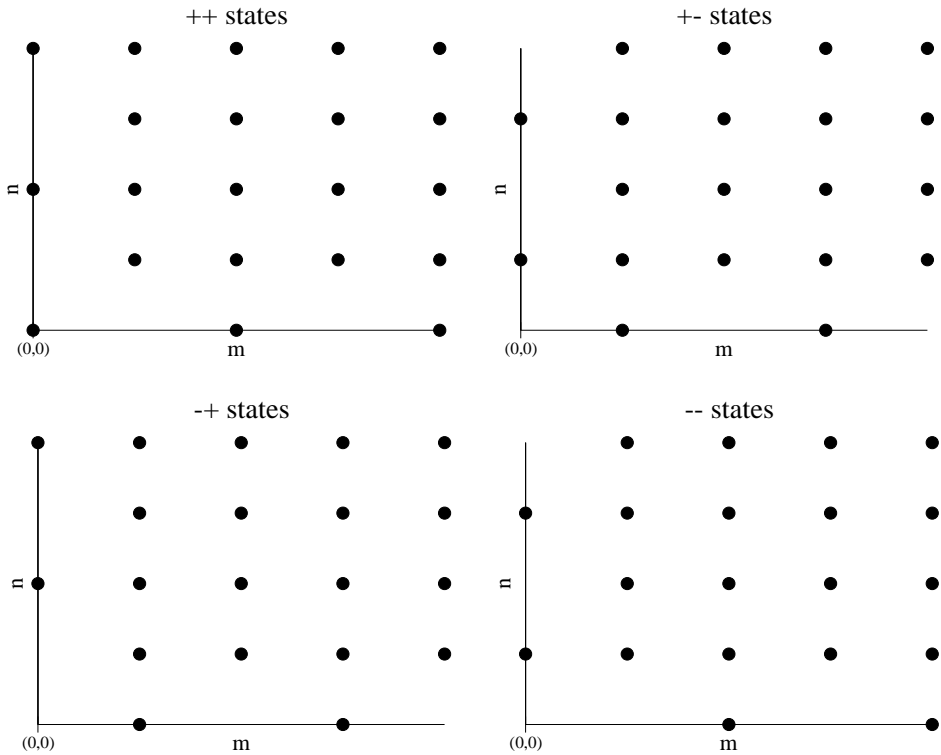

\begin{minipage}[c]{0.5\linewidth}
\centering
\includegraphics[width=5.8cm]{statespp.eps}
\vspace{15pt}
\end{minipage}
\begin{minipage}[c]{0.5\linewidth}
\centering
\includegraphics[width=5.8cm]{statespm.eps}
\vspace{15pt}
\end{minipage}
\begin{minipage}[c]{0.5\linewidth}
\centering
\includegraphics[width=5.8cm]{statesmp.eps}
\end{minipage}
\begin{minipage}[c]{0.5\linewidth}
\centering
\includegraphics[width=5.8cm]{statesmm.eps}
\end{minipage}
\caption{\label{states}\footnotesize The mode expansion in Eq.(\ref{modeexp})
gives the information about where the various parity eigenstates exist. Notice
that this figure depicts only the positive parts of the (m,n) values while for
the calculations they should be summed over both positive and negative integers.
It is clear from the figure that only the $(++)$ fields have zero modes.}
\end{figure}

The eigenstates of the parity operations are required to obey:
\beqn
\varphi_{\pm \widehat \pm} (x_\mu, -\xf, -\xs) &=& \pm  \varphi_{\pm \widehat
\pm} (x_\mu, \xf, \xs) \nonumber \\
\varphi_{\pm \widehat \pm} (x_\mu, -\xf+ \pi \Rf, \xs + \pi \Rs) &=& \widehat
\pm \varphi_{\pm \widehat \pm} (x_\mu, \xf, \xs)
\eeqn
which project out even and odd modes that can be written out as:
\beqn
&& \varphi_{\pm \widehat{\pm}}(x, \xf, \xs) = \frac{1}{4\sqrt{2 \pi \Rf \Rs}}
\nonumber \\
&&\times \sum_{m, n} \left[(\varphi^{(m,n)} \pm \varphi^{(-m,-n)}) \widehat{\pm}
(-1)^{m-n}(\varphi^{(-m,n)} \pm \varphi^{(m,-n)})  \right]\text{exp} \left[ i
\left( \frac{m\xf}{\Rf} + \frac{n \xs}{\Rs} \right) \right] \nonumber\\
\label{modeexp}
\eeqn
In the above three expressions, $\pm$ denotes states that are even/odd under the
first parity operation, and $\widehat{\pm}$ denotes states that are even/odd
under the second parity.  The massless modes come only from the $+\widehat{+}$
(hereafter denoted as ++) parity modes. The above spectrum is illustrated in
Fig. \ref{states}

\subsection{SU(5) $\rightarrow$ SU(3) $\times$ SU(2) $\times$ U(1)$_Y$}
We now introduce a Wilson line to break the symmetry down to the Standard Model.
A gauge field, $A_M \equiv \sum_a A_M^a T^a$ transforms under a gauge
transformation as follows:
\beq
A_M(x_\mu, \xf, \xs) \rightarrow U A_M(x_\mu, \xf, \xs) U^\dagger - i U
\partial_M U^\dagger
\eeq
where $T^a$ correspond to the generators of the gauge group.\footnote{This is
the remaining gauge symmetry of the supersymmetric theory in the Wess-Zumino
gauge.} Now consider a constant background gauge field along the fifth ans sixth
directions:
\beq
A_5 = \frac{1}{4\Rf} T \quad \text{and}, \quad A_6 = \frac{1}{4\Rs} T
\eeq
where $T$ is the generator (up to a constant) that breaks SU(6) down to SU(3)
$\times$ SU(3) $\times$ U(1) given by:\footnote{This constant background field
is consistent with the parity operation $A_5 \rightarrow - A_5$ with the
additional periodic gauge transformation, such that
$A_5' = U(\xf) (- A_5) U(\xf)^\dagger -i U(\xf) \partial_{\xf} U(\xf)^\dagger
\equiv A_5$ and $U(\xf) = \text{exp} \left( -i \frac{\xf}{\Rf} \frac{T}{2}
\right)$ is periodic under $\xf \rightarrow \xf + 2 \pi \Rf$ up to an element of
the center of the group SU(6) \cite{Dermisek:2002ri}.}
\begin{align}
T = \left( \begin{array}{cccccc}
 1&  &  &  &  &  \\
 & 1 &  &  &  &  \\
 &  & 1 &  &  &  \\
 &  &  & -1 &  &  \\
 &  &  &  &  -1&  \\
 &  &  &  &  & -1 \\
 \end{array} \right)
\end{align}
Note that the choice of the background gauge fields must obey some strict constraints. For example, the space group generators obey:
\begin{align}
 {\cal Z}^2 = \mathds{1}, \qquad {\cal Z'}^2 = T_6
\end{align}
The second condition implies that the action of the parity ${\cal Z'}$ is equivalent to the holonomy coming from the gauge field along the sixth direction. In addition,
\beq
{\cal Z Z' Z Z'} = T_5^{-1}
\eeq
Rewriting the above relation of the space group generators as holonomies, we get:
\beq
G({\cal Z}^2) G({\cal Z'}^2) = G(T_5^{-1})
\eeq
where we have use the fact that U(1) holonomies commute. Noting that $G(T_5^{-1})$ = $G(T_5)$, we find that the holonomies should obey the condition:
\beq
G(T_5) = G(T_6)
\label{holonomies}
\eeq
This statement tells us that the Wilson lines cannot be independent along the two extra-dimensions\footnote{We are thankful to the referee for pointing this out.}.
The presence of such a background gauge field breaks the gauge
symmetry.\footnote{This mechanism is popularly known as \textit{Hosotani}
mechanism or \textit{Wilson-line} symmetry breaking\cite{Hosotani:1983xw,
Candelas:1985en, Witten:1985xc}.} The constant background fields introduce a
holonomy equal to $W = \text{exp} \left( i \oint A_{5} d \xf + i \oint A_{6} d
\xs \right)$. This non-trivial holonomy affects the spectrum of Kaluza-Klein
states. In an
equivalent picture\cite{Hall:2001tn, Dermisek:2002ri}, the background gauge
field can be gauged away completely by choosing the proper gauge transformation, and
in this case, we find that the gauge condensate vanishes when
\beq
U(\xf) = \text{exp} \left[i \left( \frac{\xf}{\Rf} + \frac{\xs}{\Rs} \right)
\frac{T}{4} \right]
\label{eqn:gauge}
\eeq
Nevertheless, the physics remains unchanged, and we determine the change in the
KK spectrum due to the non-trivial holonomy (or Wilson-line).

Under the gauge transformation operator, Eq.(\ref{eqn:gauge}), a generic adjoint
field $\varphi$ transforms as:
\beq
\varphi'(x_\mu, \xf, \xs) = U(\xf, \xs) \varphi(x_\mu, \xf, \xs) U^\dagger(\xf,
\xs)
\eeq
which allows us to rewrite the gauge transformed wave function as
\beq
\varphi'(x_\mu, \xf, \xs) = \text{e}^{ i \left( \frac{\xf}{\Rf} +
\frac{\xs}{\Rs} \right) \frac{\irho}{4}} \varphi(x_\mu, \xf, \xs)
\label{transformed}
\eeq
where, $I_\rho$ is the eigenvalue of the generator T and $ \varphi(x_\mu,
\xf,\xs)$ is the untransformed wave function as defined in Eq.
(\ref{expansion}). The periodicity condition Eq.(\ref{periodicity}) of the
fields then becomes:
\beqn
\varphi'(x_\mu, \xf+ 2\pi \Rf, \xs) &=& P' \varphi'(x_\mu, \xf, \xs) P'^\dagger
\equiv e^{i \frac{\pi}{2} I_\rho} \varphi'(x_\mu, \xf, \xs)
\nonumber \\
\varphi'(x_\mu, \xf, \xs + 2\pi \Rs) &=& P' \varphi'(x_\mu, \xf, \xs) P'^\dagger
= e^{i \frac{\pi}{2} I_\rho} \varphi'(x_\mu, \xf, \xs)
\eeqn
where $P' \equiv \text{exp} \left(i \frac{\pi}{2} T \right) = \text{diag}(i, i,
i, -i, -i, -i)$. The above equation reflects the constraints on the Wilson lines
that was demonstrated in Eq. (\ref{holonomies}). In addition, now we have
re-expressed the Wilson line as a parity operation
that breaks SU(6) down to SU(3) $\times$ SU(3) $\times$ U(1). Under the combined
parity operations on the manifold and the non-vanishing background fields along
the fifth and sixth directions, we have achieved gauge symmetry breaking of the SU(6) group
to [SU(3) $\times$ SU(2) $\times$ U(1)$_Y$] $\times$ U(1)$_X$. The only choice we had here was a combination
of local and non-local GUT breaking. It is possible to have a purely non local GUT
breaking if one started with an SU(5) gauge theory on the same orbifold and and
chose the second parity to break the SU(5) down to SU(3) $\times$ SU(2) $\times$
U(1) \cite{Hebecker:2003we}.

We still have to calculate how the mass spectrum changes as a result of the
holonomy due to the gauge field. This can be easily done by looking at the
transformed wave function in Eq.(\ref{transformed}) and calculating the
eigenvalues \irho\ of the generator T. The eigenvalues $I_\rho$ can be
determined by calculating the commutator $\left[T, \varphi \right] $ since
$\varphi$ is in the adjoint representation, of the form:
\begin{align}
\varphi =  &\left( \begin{array}{ccccc|ccc|c}
&&&&&&&&  \\
&&(8,1)_0&&&&(3, \bar 2)_{-5/3}&&(3,1)_{-2/3}  \\
&&&&&&&&  \\ \hline
&&&&&&&&  \\
&&(\bar{3}, 2)_{5/3}&&&&(1,3)_0&&(1,2)_1 \\ \hline
&&(\bar{3},1)_{2/3}&&&&(1, \bar 2)_{-1} &&(1,1)_0 \\
 \end{array} \right)& \nonumber \\
= &\left( \begin{array}{ccccc|ccc|c}
&&&&&&&&  \\
&&g&&&&X&&T  \\
&&&&&&&&  \\ \hline
&&&&&&&&  \\
&&\bar{X}&&&&w&&H_u \\ \hline
&&\bar{T}&&&&H_d&&b \\
 \end{array} \right)&
\end{align}
The first line in the above expression shows the quantum numbers of the the
different blocks that the adjoint field gets broken into after the orbifold
projection and holonomy. We name them appropriately, so that they can be
associated with the fields that remain massless in the low energy theory, like
the gauge bosons, $g, w, b$ and the Higgs doublets, $H_u, H_d$; and the fields
that obtain mass and do not appear in the low energy spectrum like the Higgs
triplets $T, \bar{T}$ and states with exotic quantum numbers $X, \bar{X}$. The
commutator of the generator $T$ with this quantity is calculated and the
eigenvalues of are summarized in Table \ref{Irho}.

\begin{table}
\begin{center}
\begin{tabular}{|c|c|c|c|c|c|c|c|c|c|}
\hline
& $g$ & $w$ & $b$ & $X$ & $\bar{X}$ & $T$ & $\bar{T}$& $H_u$ & $H_d$ \\ \hline
$I_\rho$ & 0& 0&0&2&-2&2&-2&0&0\\ \hline
\end{tabular}             \end{center}
\caption{Eigenvalues $I_\rho$ of the generator T acting on the various fields
(labelled by $\rho$) in
the model.}
\label{Irho}
\end{table}

Eventually, we see that the masses of the states in the KK tower are given by
\beq
M_{(m,n), \rho}^2  = \frac{(m+\frac{I_\rho}{4})^2}{\Rf^2} +
\frac{(n+\frac{I_\rho}{4})^2}{\Rs^2}
\eeq
The massless states are those which are even under both parities and have zero
eigenvalue under the holonomy. These turn out to be only the standard model
gauge bosons and the Higgs doublets, $H_u, H_d$ coming from the chiral adjoint
$\Sigma_6$. Finally, we also note that at the two fixed points, $F_1$ and $F_2$
which are located at $(0,0)$ and $(0,
\pi \Rs)$, there is only an SU(5) whereas the bulk has an SU(6). The three
families of quarks and leptons are also assumed to sit at these singularities
coming in 3 ($\bf 10_F + \bar 5_F$) representations. The Yukawa couplings are
also assumed to be localized at these fixed points. They require superpotential
terms of the form $\bf 10_F \ 10_F \ 5_{\Sigma_6} + 10_F \ \bar 5_F \ \bar
5_{\Sigma_6}$ where the indices are contracted in an obvious way.
The SU(5) relation $\lambda_b = \lambda_\tau$ works for the third family but not
for the first two.  It is possible
that interaction with matter in the bulk could help with this issue, but this is
beyond the
scope of this paper.

\subsection{Proton decay}

Dimension 6 operators for proton decay are suppressed by the inverse power
squared of the smallest compactification scale. We will see that this is near
the 4D GUT scale and thus the proton lifetime is completely consistent with the
experimental bounds. On the other hand, dimension 5 operators for proton decay
are only suppressed by the inverse power of the compactification scale. However,
if we assume that quarks and leptons only couple to the chiral adjoints
containing the Higgs fields,  there are no dimension 5 operators for proton
decay generated when integrating out the color triplet Higgs fields.  This can
be attributed to an unbroken $\mathbb{Z}_4^R$ symmetry \cite{Lee:2010gv} where
the superpotential has charge 2, families have charge 1, $\{ \Sigma_{5, 6}, \ 6,
\ \bar 6 \}$ have charge 0, and $\{ S, \ \Phi \}$ have charge 2.

\section{Threshold Corrections}
\label{unification}
4D SUSY GUTs require extra states to contribute a small amount of threshold
corrections at the GUT scale in order to concur with low energy measurements.
Conventionally, this quantity of GUT scale threshold corrections (defined at the
4D GUT scale) is defined as:
\beq
\epsilon_3 = \frac{\alpha_{3} -\alpha_{GUT}}{\alpha_{GUT}}
\label{4d} .
\eeq
The running coupling constants in the 4D MSSM can be summarized by:
\beq
 \alpha_{i}^{-1} (Q) = \alpha_{GUT}^{-1} + \frac{b_{i}}{2\pi} log
\frac{M_{GUT}}{Q} - \alpha_{GUT}^{-1} \frac{\epsilon_3}{(1 + \epsilon_3)}
\delta_{i3} \label{4drg}
\eeq
where $\delta_{i3}$ denotes that the term appears only for i=3 (the coupling
$\alpha_3$). The exact amount of threshold corrections required from the extra
states is usually model dependent, but they have to be around a few percent
level. For the most popular scenarios of MSSM with unified gaugino masses, this
number turns out to be about -3\%. We would like to calculate the effect of the
Kaluza-Klein (KK) tower of infinite states to the running of coupling constants
in the orbifold model that we have just constructed. These additional
contributions to the running of the coupling constants from KK modes can be
written as:\footnote{We have followed the analysis
of Ref. \cite{Ghilencea:2003kt} in what follows.  The details can be found in
the Appendix \ref{kki}.}
\beq
\frac{4 \pi}{g_{i}^{2}(\mu)} = \frac{4\pi}{g^{2}(\Lambda)}  + \sum_{\rho}
\Omega_{i, \rho} (\mu)
\label{org}
\eeq
where
\beq
\Omega_{i, \rho} (\mu) \equiv \frac{1}{4 \pi} \sum_{(m,n) \in Z} \beta_{i, \rho}
\int_{\xi}^{\infty} \frac{dt}{t} e^{- \pi t \frac{M_{(m,n), \rho}^2 }{\mu^2}}
e^{-\pi \chi t}
\label{KKintegral}
\eeq
includes one-loop corrections from both massive and massless states in the
theory. $\xi$ is the ultraviolet (UV) regulator introduced since the integral is
UV-divergent. $\chi$ is an infrared (IR) regulator introduced since the above
quantity diverges for the special case when there are massless states in the KK
tower. The corrections come from each state $\rho$ that appears in the spectrum,
with an associated beta-function coefficient, $\beta_{i,\rho}$, summarized in
Table \ref{beta} and mass, $M_{(m,n), \rho}^2 $, as calculated in the previous
section:
\beq
M_{(m,n), \rho}^2  = \frac{(m+\frac{I_\rho}{4})^2}{\Rf^2} +
\frac{(n+\frac{I_\rho}{4})^2}{\Rs^2}
\label{mass}
\eeq
We evaluate the expression in Eq. (\ref{KKintegral}) in three different regions
on the m-n plane shown in Fig \ref{states} and then sum up the contributions to
find the total corrections to the couplings.

\subsection{States at m = 0 and n = 0}
In this case, the contribution to the threshold corrections is:
\beq
\Omega_{i, \rho}^{00} (\mu) = \frac{1}{4 \pi} \beta_{i, \rho}
\int_{\xi}^{\infty} \frac{dt}{t} e^{- \pi t \frac{M_{(0,0), \rho}^2 }{\mu^2}}
e^{-\pi \chi t}
\eeq
We saw earlier that the only states at the m=0, n=0 point are the (++) modes.
The (++) modes come from the N=1 SUSY vector fields $g,w,b, X, \bar{X},$ and
chiral adjoint fields  $T, \bar{T}, H_u, H_d$. The beta-function coefficients
for these states are summarized in Table \ref{beta}. Using the results from
Appendix, we find:
\begin{small}
\beqn
\Omega_i^{00} = \frac{b^{++}_i (\irho = 0)}{4 \pi} \Gamma \left[0, \pi \xi \chi
\right] + \frac{b^{++}_i (\irho = 2)}{4 \pi} \Gamma \left[0, \pi \xi \left(
\frac{1}{4 \mu^2 \Rf^2} + \frac{1}{4 \mu^2 \Rs^2} +  \chi \right) \right]
\eeqn
\end{small}

\begin{table}[ht]
\begin{center}
\begin{small}
\begin{tabular}{|c|c|c|c|c|c|c|c|c|c|}
\hline
Quantum Number& Name & Type & $b_1$ & $b_2$ & $b_3$ & Type & $b_1$ & $b_2$ &
$b_3$\\
\hline \hline
(\textbf{8},1)$_0$ & $g$ & C & 0 & 0 & 3 & V & 0 & 0 & -9  \\
(1,\textbf{3})$_0$ & $ w $ & C & 0 & 2 & 0 & V & 0 & -6 & 0  \\
\hline
(\textbf{3},\textbf{2})$_{\pm 5/3}$ & $X, \bar{X}$ & C & 5/2 & 3/2 & 1 & V &
-15/2 & -9/2 & -3 \\
\hline
(\textbf{3},1)$_{\pm 2/3}$ & $T, \bar{T}$ & C & 1/5 & 0 & 1/2 & V & -3/5 & 0 &
-3/2 \\
\hline
(1,\textbf{2})$_{\pm 1}$ & $H_u, H_d$&C & 3/10 & 1/2 & 0 & V & -9/10 & -3/2 & 0
\\ \hline
\end{tabular}
\end{small}
\caption{Nomenclature, Quantum numbers, and beta-function coefficients for the
various states in the spectrum. \label{beta}}
\end{center}
\end{table}
\subsection{m axis, n = 0}
Figure \ref{states} shows that the $(++)$ and $(--)$ states live only at even n
whereas $(+-)$ and $(-+)$ states live at odd n. The absence of states at certain
n has to be accounted for while evaluating the integral. The details of
evaluating the odd and even integrals are
explicitly presented in Appendix \ref{kki} and the result is:
 \begin{small}
\beqn
\Omega_i^{m0} &=& \frac{b_i^{(++)} (\irho = 0)}{4 \pi} {\cal R}_1^E \left[ \xi
\nu_1, 0, \frac{\delta_1}{\nu_1} \right] + \frac{b_i^{(++)} (\irho = 2)}{4 \pi}
{\cal R}_1^E \left[ \xi \nu_1, 1/2, \frac{\delta_1}{\nu_1} \right] \nonumber \\
&+& \frac{b_i^{(+-)} (\irho = 0)}{4 \pi} {\cal R}_1^O \left[ \xi \nu_1, 0,
\frac{\delta_1}{\nu_1} \right] + \frac{b_i^{(+-)} (\irho = 2)}{4 \pi} {\cal
R}_1^O
\left[ \xi \nu_1, 1/2, \frac{\delta_1}{\nu_1} \right] \nonumber \\
&+& \frac{b_i^{(-+)} (\irho = 0)}{4 \pi} {\cal R}_1^O \left[ \xi \nu_1, 0,
\frac{\delta_1}{\nu_1} \right] + \frac{b_i^{(-+)} (\irho = 2)}{4 \pi} {\cal
R}_1^O
\left[ \xi \nu_1, 1/2, \frac{\delta_1}{\nu_1} \right] \nonumber \\
&+& \frac{b_i^{(--)} (\irho = 0)}{4 \pi} {\cal R}_1^E \left[ \xi \nu_1, 0,
\frac{\delta_1}{\nu_1} \right] + \frac{b_i^{(--)} (\irho = 2)}{4 \pi} {\cal
R}_1^E
\left[ \xi \nu_1, 1/2, \frac{\delta_1}{\nu_1} \right] \nonumber \\
\Omega_i^{m0} &=& \left( \frac{b_i^{(++)} (\irho = 0)}{4 \pi} +\frac{b_i^{(--)}
(\irho = 0)}{4 \pi} \right)  {\cal R}_1 \left[4 \xi \nu_1, 0,  \frac{\chi}{4
\nu_1} \right] \nonumber \\
&+& \left(\frac{b_i^{(+-)} (\irho = 0)}{4 \pi} + \frac{b_i^{(-+)} (\irho = 0)}{4
\pi} \right) \left( {\cal R}_1 \left[ 4 \xi \nu_1, \frac{1}{2},
\frac{\chi}{4
\nu_1} \right] + \Gamma \left[0, \pi \xi \left(\nu_1+ \chi \right) \right]
\right) \nonumber \\
&+& \left( \frac{b_i^{(+-)} (\irho = 2)}{4 \pi} + \frac{b_i^{(-+)} (\irho =
2)}{4 \pi}\right) \Gamma \left[0, \pi \xi  \left(\frac{\nu_1}{4}
+\frac{\nu_2}{4} + \chi \right)
\right]
\eeqn
\end{small}
where, $\nu_1 = \frac{1}{\mu^2 \Rf^2}$, $\nu_2 = \frac{1}{\mu^2 \Rs^2}$, and
$\delta_1 = \frac{\rho_2}{\mu^2 \Rs^2} + \chi$

The function ${\cal R}_1$ is also defined in Appendix \ref{kki}. In simplifying
the above expression, we have also used the fact that when we have complete N=4
SUSY in 4D,
the beta-function coefficients sum up to zero.
\beq
b_i^{(++)} + b_i^{(+-)} + b_i^{(-+)} + b_i^{(--)} = 0
\eeq
for all i\footnote{We have complete N=4 SUSY in 4D when we have one vector
multiplet and three chiral
multiplets. In terms of the N=1 fields in 4D, the beta-function coefficients are
given by:
\beq
b_G = 3 C_2 (G) - N_{\text{chiral}} T{(R)}
\eeq}.

\subsection{n axis, m = 0}
Along this axis, the calculation is similar to the previous case in the sense
that the states
exist only at certain n. The $(++)$ and $(-+)$ states live only at even n
whereas $(+-)$ and
$(--)$ states live at odd n. Again, using the relations in Appendix \ref{kki}
and evaluating the
integrals, we get:
\begin{small}
\beqn
\Omega_i^{0n} &=& \frac{b_i^{(++)} (\irho = 0)}{4 \pi} {\cal R}_1^E \left[ \xi
\nu_2, 0, \frac{\delta_2}{\nu_2} \right] + \frac{b_i^{(++)} (\irho = 2)}{4 \pi}
{\cal R}_1^E \left[ \xi \nu_2, 1/2, \frac{\delta_2}{\nu_2} \right] \nonumber \\
&+& \frac{b_i^{(+-)} (\irho = 0)}{4 \pi} {\cal R}_1^O \left[ \xi \nu_2, 0,
\frac{\delta_2}{\nu_2} \right] + \frac{b_i^{(+-)} (\irho = 2)}{4 \pi} {\cal
R}_1^O \left[ \xi \nu_2, 1/2, \frac{\delta_2}{\nu_2} \right] \nonumber \\
&+& \frac{b_i^{(-+)} (\irho = 0)}{4 \pi} {\cal R}_1^E \left[ \xi \nu_2, 0,
\frac{\delta_2}{\nu_2} \right] + \frac{b_i^{(-+)} (\irho = 2)}{4 \pi} {\cal
R}_1^E \left[ \xi \nu_2, 1/2, \frac{\delta_2}{\nu_2} \right] \nonumber \\
&+& \frac{b_i^{(--)} (\irho = 0)}{4 \pi} {\cal R}_1^O \left[ \xi \nu_2, 0,
\frac{\delta_2}{\nu_2} \right] + \frac{b_i^{(--)} (\irho = 2)}{4 \pi} {\cal
R}_1^O \left[ \xi \nu_2, 1/2, \frac{\delta_2}{\nu_2} \right] \nonumber \\
\Omega_i^{0n} &=& \left( \frac{b_i^{(++)} (\irho = 0)}{4 \pi} +\frac{b_i^{(-+)}
(\irho = 0)}{4 \pi} \right)  {\cal R}_1 \left[4 \xi \nu_2, 0,  \frac{\chi}{4
\nu_2} \right] \nonumber \\ &+& \left(\frac{b_i^{(+-)} (\irho = 0)}{4 \pi} +
\frac{b_i^{(--)} (\irho = 0)}{4 \pi} \right) \left( {\cal R}_1 \left[ 4 \xi
\nu_2, \frac{1}{2}, \frac{\chi}{4 \nu_2} \right] + \Gamma \left[0, \pi \xi
\left(\nu_2+ \chi \right) \right] \right) \nonumber \\
&+& \left( \frac{b_i^{(+-)} (\irho = 2)}{4 \pi} + \frac{b_i^{(--)} (\irho =
2)}{4 \pi}\right) \Gamma \left[0, \pi \xi \left(\frac{\nu_2}{4}+
\frac{\nu_1}{4} + \chi \right) \right]
\eeqn
\end{small}
where, $\nu_1 = \frac{1}{\mu^2 \Rf^2}$, $\nu_2 = \frac{1}{\mu^2 \Rs^2}$, and
$\delta_2 = \frac{\rho_1}{\mu^2 \Rf^2} + \chi$ as defined in Appendix \ref{kki}.

\subsection{Off the axes}
\label{mnne0}
This case turns out to be rather simple since all the parity eigenstates live at
all (m,n) $\ne\ 0$. This includes one vector and three chiral adjoint multiplets
for every
state and they form complete N=4 supersymmetry. Thus these excited
KK modes do not contribute anything to the running of the coupling constants.

\subsection{Putting it all together}
The contribution from the four individual cases can be put together with the
appropriate beta-function coefficients. In the limit that the regulators can be
set to zero, they can be combined with the mass scale $\mu$ and replaced by
their relevant UV and IR scales.
\begin{align}
 Q^2 \equiv \pi e^\gamma \chi \mu^2 \Big|_{\chi \rightarrow 0} \qquad \Lambda^2
\equiv \frac{\mu^2}{\xi}\Big|_{\xi \rightarrow 0}
\end{align}
The functions $\Gamma$ and ${\cal R}_1$ in these limits simplify and these
simplified expressions are summarized in Appendix \ref{function_limits}. The
final expression for the threshold corrections at the scale Q coming all the KK
states that
exist in the system are given by:
\begin{small}
\beqn
\Omega_{i}(Q) &=& \frac{b_i^{++} (\irho = 0)}{4 \pi} \text{ln}
\frac{\Lambda^2}{Q^2} + \left(\frac{b_i^{+-} (\irho = 0) + b_i^{-+} (\irho =
0)}{4 \pi}\right) \text{ln} \left[\frac{\pi \Lambda}{2 M_5}
\right]^2  \nonumber \\
&+& \left(\frac{b_i^{+-} (\irho = 0) + b_i^{--} (\irho = 0)}{4 \pi}\right)
\text{ln} \left[\frac{\pi \Lambda}{2 M_6} \right]^2 \nonumber \\
&+& \frac{b_i^{+-} (\irho = 2)} {4 \pi}
\text{ln} \left[ \frac{4 \Lambda^2}{ M_5^2+
M_6^2}\right]
\label{corrections}
\eeqn
\end{small}
where the scales $M_i, i = 5,6$ are rescaled compactification scales, i.e.  $M_i
= \frac{\sqrt{\pi e^{\gamma}}}{R_i}$. Note that to arrive at this result, we
have used the spectrum in Fig. \ref{states} with mass eigenvalues as shown in
Eq.(\ref{mass}). The important feature of this expression is that it tells us
that there are no power-law corrections to the couplings at any scale. This is
unlike generic scenarios of a (4+$\delta$)D model with $\delta$ compactified
dimensions, where the couplings receive power-law corrections proportional to
$\left(\frac{\Lambda}{M_C}\right)^{\delta}$ where $M_C$ is the smallest
compactification scale. Therefore, we should have expected quadratic
corrections to the couplings in the 6D model considered here. It turns out that
the quadratic corrections vanish due to the initial N=4 SUSY. This feature was
also observed in Ref. \cite{Hall:2001zb} where an SU(6) theory was studied with
N=2 SUSY in 6D. The model discussed in\cite{Hall:2001zb}, however had an
effective 5D limit. Hence there were additional linear corrections to the
couplings. In the model discussed here, the compactification takes the 6D theory
directly down to 4D and hence we find only logarithmic corrections to the
couplings.
\begin{table}
\begin{center}
\begin{tabular}{|c|c|c|}
\hline
 Coefficients & $(b_1, b_2, b_3)$ \\
\hline
$b_i^{++} (\irho = 0)$ & $(\frac{33}{5}, 1, -3)$  \\
$ b_i^{+-} (\irho = 0) + b_i^{-+} (\irho = 0)$ &$(-\frac{6}{5}, 2, 6)$\\
$b_i^{+-} (\irho = 0) + b_i^{--} (\irho = 0) $ & $(\frac{6}{5}, 6, 6)$ \\
$b_i^{+-} (\irho = 2)$ &$(\frac{27}{5},3,3)$ \\
\hline
\end{tabular}
\caption{Beta-function coefficients relevant for Eq. (\ref{corrections})}
   \label{finalbeta}        \end{center}
\end{table}

\section{Results \& Discussion}
\label{results}
We now compare the result we obtained in Eq. (\ref{corrections}) from the 6D
orbifold to the gauge couplings of the low energy 4D MSSM and determine if the
spectrum obtained can account for the correct amount of GUT-scale threshold
corrections as required by the standard scenarios of MSSM, about -3 \%, when the
4D GUT-scale, \mgut\  is
around 3 $\times 10^{16}$ GeV. If the low energy limit of the orbifold
construction is the same as the MSSM, then
at energies below the smallest of the compactification scales, $M_C$, the
couplings should be the same for both theories. Above $M_C$ new states appear in
the orbifold GUT and then, the running of the couplings differ in the two
theories. In the 4D MSSM, it is believed that the couplings unify at a grand
unification scale, with small
corrections from states near that scale that spoil precision unification. If
$M_C$ happens to be close to the 4D GUT scale and we obtain the appropriate
threshold corrections, then we have an alternate understanding of \mgut. The 4D
GUT scale in this case is just a fictitious scale obtained by running the
couplings from the weak scale up. However it can now be identified with the
compactification scale, where all the new physics arises. At the same time, the
real unification naturally happens at the cut-off scale. This scale would be
identified with the string scale, assuming the underlying theory of an orbifold
GUT is string theory.

At the lowest compactification scale (largest compactification radius), we have
6D orbifold and the 4D MSSM, respectively:
\beqn
 \alpha_{i}^{-1} (Q) &=& \alpha^{-1} (\Lambda) +
\sum_{\rho} \Omega_{i, \rho} (Q) \nonumber \\
 \alpha_{i}^{-1} (Q) &=& \alpha_{GUT}^{-1} + \frac{b_{i}}{2\pi} log
\frac{M_{GUT}}{Q} - \alpha_{GUT}^{-1} \frac{\epsilon_3}{(1 + \epsilon_3)}
\delta_{i3} \nonumber
\eeqn
We have 3 sets of equations, one for each coupling of SU(3) $\times$ SU(2)
$\times$
U(1)$_Y$ and four unknowns: $\Lambda$, $M_5$, $M_6$, and $\alpha(\Lambda)$, the
unified coupling constant
 of the orbifold theory, given \mgut\ and $\epsilon_3$ from the 4D MSSM. We find
that we can uniquely solve for
$M_5$ and $M_6$ in terms of \mgut\ and $\epsilon_3$ and we obtain a curve in the
$\alpha - \Lambda$
plane. The details of the solution are elaborated in Appendix
\ref{analyticalsolution} and we summarize the solutions
obtained:
\beqn
M_5 &=& \left( m(\epsilon_3)^{({\cal G- H})/2} (m(\epsilon_3)+1)^{{\cal H}/2}
e^{{\cal I}/2}\right) \mgut \nonumber \\
M_6 &=& \left( m(\epsilon_3)^{({\cal G- H}-1)/2} (m(\epsilon_3)+1)^{{\cal H}/2}
e^{{\cal I}/2}\right) \mgut \nonumber \\
\alpha^{-1}(\Lambda) &=& - \frac{3}{\pi} \text{ln} \frac{\Lambda^2}{M_{GUT}^2} +
\frac{3}{\pi} \text{ln} \left(m(\epsilon_3)^{({\cal G- H})}
(m(\epsilon_3)+1)^{{\cal H}}  e^{{\cal I}} \right) \nonumber \\ &+& \text{ln}
\left(m(\epsilon_3)^{({\cal L- M})} (m(\epsilon_3)+1)^{{\cal M}}  e^{{\cal N}}
\right)
\label{al}
\eeqn
\begin{figure}[ht!]
\begin{center}
\includegraphics[width=8cm]{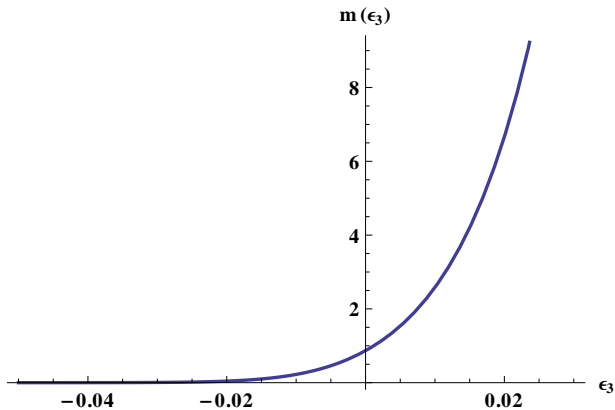}\end{center}
\caption{\label{me3}The figure shows the dependence of $m =
\left(\frac{M_5}{M_6}\right)^2$ on $\epsilon_3$. The statement that MSSM
requires small threshold corrections at the GUT scale translates to anisotropic
compactification.}
\end{figure}

The coefficients ${\cal G, H, I}$ and ${\cal N}$ are given in Table \ref{cal} in Appendix
\ref{analyticalsolution}. To analyze the GUT scale threshold corrections, we
fix \iag\ to be 24 in all further calculations. Benchmark points are shown in
Table \ref{benchmark}. The ratio of $M_5$ and $M_6$ = $m$, depends only on $\epsilon_3$
and is shown in Fig \ref{me3}. The value of m sets the hierarchy between the two compactification scales, $M_5$
and $M_6$. We analyzed the particle spectrum at intermediate energies in the cases when
(i) $M_5 \ \ll \ M_6$ (ii)  $M_6 \ \ll \ M_5$ and (iii)  $M_5 \ = \ M_6$ to determine the scale associated
with the unication of SU(3) $\times$ SU(2) $\times$ U(1)$_Y$ gauge groups. Also, to determine if the SU(6) was broken down to a subgroup at these intermediate scales, reflecting the two step GUT breaking
procedure that we employed. We find two unification scales - the SM gauge group unify to an SU(3) $\times$ SU(3) $\times$ U(1) at the scale $M_5$ in all the above three cases. Then further at the scale $\sqrt{M_5^2 + M_6^2}$ there is
another unification scale associated with SU(3) $\times$ SU(3) unification to the SU(6) GUT.
On the other hand,  we do not find a scale associated with the breaking to SU(5)
$\times$ U(1)$_X$ which is a typical feature of local GUT breaking as noted in \cite{Trapletti:2006xv}.

\begin{figure}[H]
\begin{center}
\vspace{1cm}
 \includegraphics[height=7.4cm, width=9.5cm]{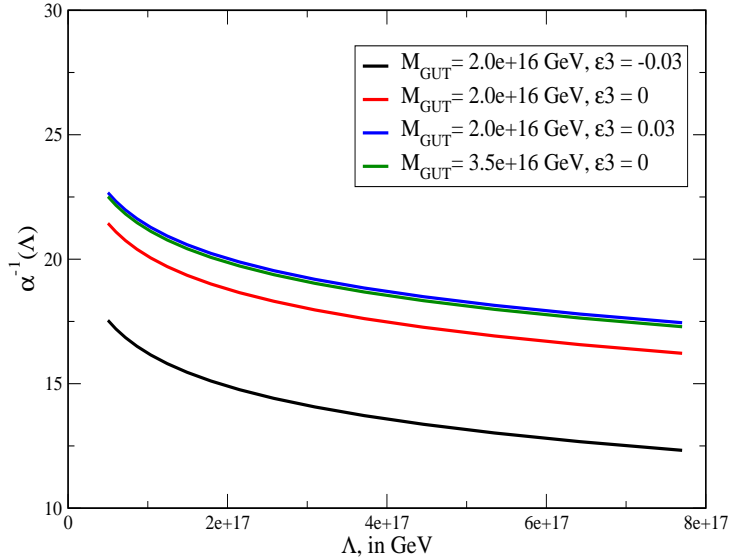}\\
\caption{\label{alpha}Once $M_5$ and $M_6$ are solved for uniquely, we are left
with a curve in the $\alpha^{-1}-\Lambda$ plane, as expressed in Eq. (\ref{al}).
The unified coupling at the cut-off scale is in the perturbative regime.}
\end{center}
\end{figure}

It is also interesting to note that the standard scenarios of the MSSM can be
embedded in an isotropic or anisotropic orbifold. We find that in the
anisotropic as well as isotropic ($M_5 \sim M_6$) cases, the lowest
compactification scale is around the 4D GUT scale, making it possible to connect
the compactification scale and the 4D GUT scale. For three benchmark points, the
curve in the $\alpha^{-1} (\Lambda) - \Lambda$ plane, from Eq. (\ref{al}) is
shown in Fig \ref{alpha}.   Finally we note that the values of $\alpha^{-1}
(\Lambda), \Lambda$ are not consistent with perturbative heterotic string
boundary conditions.  In particular, since $\alpha$ depends only of the
logarithm of $\Lambda$, it is not possible to
embed this orbifold GUT into the weakly coupled regime of the heterotic string,
where value of the GUT coupling constant at the string scale is given
by\cite{Dundee:2008tr}:
\beq
\alpha^{-1}(\Lambda = M_{string}) = \frac{1}{8} \left(\frac{M_{PL}}{M_{string}}
\right)^2
\eeq

\begin{table}
\centering
 \begin{tabular}{|c|c|c|c|c|c|}
\hline
  & $\epsilon_3$  & $M_5$ & $M_6$ & $\Lambda$ & $\alpha^{-1} (\Lambda)$ \\
\hline
Point 1 & -3.0\%&  $0.174 \times 10^{16}$& $2.08 \times 10^{16}$ &  $6.0 \times
10^{17}$ & 13.57 \\ \hline
Point 2 & 0.0 \% &$3.39 \times 10^{16}$& $ 3.64 \times 10^{16}$ &  $6.0 \times
10^{17}$ & 17.47 \\ \hline
Point 3 & +3.0 \% &$1.37 \times 10^{17}$& $ 3.44 \times 10^{16}$ &  $6.0 \times
10^{17}$ & 18.70 \\
\hline
 \end{tabular}
\caption{\label{benchmark} The table shows a benchmark point for choice 1 and
choice 2. We fix \iag\ to be 24 and \mgut\ to be $3 \times 10^{16}$ GeV for both
the points. The smallest compactification scale is naturally of the order of the
4D GUT scale. All scales are in GeV.}
\end{table}

\section{Summary}
In this work, we discussed a supersymmetric SU(6) gauge theory on an orbifold
with the topology of a real projective plane.
The compact space was obtained in two steps by orbifolding a rotation and a
freely-acting roto-translation.
In the process, the gauge symmetry was broken down from SU(6) to SU(5) $\times$
U(1)$_X$ and the N=4 SUSY was reduced
to N=2. To further break the SU(5) down to the Standard Model, we introduced a
non-zero Wilson line along the
fifth and sixth directions. This helped to eliminate the unwanted light states like the
Higgs triplets and to break N=2 to N=1 SUSY.

We calculated the Kaluza Klein spectrum of states coming from this orbifolding
and also calculated the threshold corrections coming from these states at the 4D
grand unification scale. We find that the threshold corrections coming from the
KK states due to compactification on an orbifold with the topology of $RP^2$ are
at the percent level allowing for realistic 4D MSSM. The solutions allow for
threshold corrections to be between $\{-3\%, +2\%\}$ allowing for the standard
universal gaugino mass scenario like CMSSM or the non-universal gaugino mass
scenarios (especially lighter gluinos as discussed in \cite{Raby:2009sf,
Anandakrishnan:2011zn}). There have been previous calculations of threshold
corrections in orbifold GUT models on various orbifolds with local and non-local
GUT breaking. We have already pointed out that unlike in other scenarios we do
not get power law running of couplings above the compactification scale due to
the large N=2 in 6D. The advantage of not having such large power-law
corrections is that we do not lose any predictability due to UV scale physics.
We should point out that in the work of Trapletti \cite{Trapletti:2006xv}, the
author considered a non-local GUT breaking and concluded that the running of
couplings stops precisely above the compactification scale. We however find that there are small
finite threshold corrections at all scales.

Our analysis was a bottom-up approach studying the phenomenology of models on an
orbifold with the topology of a
projective plane. It would be interesting to explore the possibility of
embedding these orbifold GUTs into a more
fundamental theory, like string theory. On the other hand, it would be equally
interesting to study low energy
features like SUSY breaking and spectra. Finally, since the compactification
scale is naturally around the 4D GUT scale or larger, one does not have to
worry about proton decay from dimension 6 operators. Moreover, proton decay from
dimension 5 operators vanishes due to a discrete R symmetry.

\section{Acknowledgments}
We would like to thank Michael Ratz for useful discussions. The authors
acknowledge partial support from DOE grant
DOE/ER/01545-893. AA would also like to thank Konstantin Bobkov and Ben Dundee
for their helpful insights.

\newpage
\appendix
\section{Kaluza-Klein Integrals}
\label{kki}
In order to compute the threshold corrections coming from an infinite tower of
Kaluza-Klein states, we would like to evaluate the following integral (See
Eq.(\ref{KKintegral})):
\beq
\sum_{(m,n) \in Z} \int_{\xi}^{\infty} \frac{dt}{t} e^{- \pi t
\frac{M_{(m,n)}^2}{\mu^2}} e^{-\pi \chi t}
\eeq
where, $\chi$ and $\xi$ are IR and UV regulators, and the $M_{(m,n)}$ are the
masses of the $(m,n)^{\text{th}}$ KK mode. In the presence of Wilson lines they
are given by:
\beq
M_{(m,n)}^2 = \frac{(m + \rho_1)^2}{\Rf^2} + \frac{(n + \rho_2)^2}{\Rs^2}
\eeq
where in the scenario that we have, $\rho_1$ can be either 0 or 2 and $\rho_2$
is always 0. In general, we can solve the integral following Ghilencea
\cite{Ghilencea:2003kt} who evaluates the integral for the cases of one and two
extra-dimensions. Again, in the current scenario that we have, we find that we
only need to evaluate this integral in its one-dimensional limit. We follow
Ghilencea and evaluate a 1-dimensional Kaluza-Klein integral of the form:
\beq
{\cal R}_{1} \left[\xi, \rho, \delta \right] = \sum_{m \in Z}^{'}
\int_{\xi}^{\infty} \frac{dt}{t} e^{- \pi t \left[ (m + \rho)^2 + \delta \right]
}
\eeq
where the prime over the summation in the second term represents that m $\ne$ 0,
but runs over all other integer values.

We can make use of the Poisson re-summation formula:
\beq
\sum_{n \in Z} e^{-\pi A (n + \sigma)^2} = \frac{1}{\sqrt{A}}\sum_{\tilde{n} \in
Z} e^{-\pi A^{-1} \tilde{n}^2 + 2 i \pi \tilde{n} \sigma}
\eeq
to evaluate this integral. We have,
\beqn
{\cal R}_1 \left[\xi, \rho, \delta \right] &=& \int_\xi^{\infty} \frac{dt}{t}
\left[-e^{-\pi t \rho^2} +\sum_m e^{-\pi t (m+\rho)^2} \right] e^{-\pi \delta t}
\nonumber \\
&=& \int_\xi^\infty \frac{dt}{t} \left[-e^{-\pi t \rho^2} + \frac{1}{\sqrt{t}} +
\frac{1}{\sqrt{t}}\sum_m^{'} e^{-\pi m^2/t + 2i\pi m \rho} \right] e^{-\pi
\delta t} \nonumber \\
&=& - \Gamma \left[0, \pi \xi (\delta + \rho^2) \right] + \frac{2 e^{-\pi \delta
\xi}}{\sqrt{\xi}} + 2 \pi \sqrt{\delta} \text{Erf} \left[\sqrt{\pi \delta
\xi}\right] \nonumber \\ &-& \text{log} |2\ \text{sin} \pi (\rho + i
\sqrt{\delta})|^{2}
\eeqn
where, it has been assumed that $\xi \ll 1$ while evaluating the integral
$\int_\xi^\infty \frac{dt}{t} e^{-\pi m^2/t - \pi \delta t} $ and
Ghilencea\cite{Ghilencea:2003kt} shows that the error by doing so vanishes when
$\xi$ is small.

We summarize the result of this integral in various useful limits:
\begin{itemize}
\item (m,n) = (0,0)
\beqn
\label{00}
\int_{\xi}^{\infty} \frac{dt}{t} e^{- \pi t \frac{\rho_1^2}{\Rf^2 \mu^2}}
e^{-\pi \chi t} &=& \int_{\xi}^{\infty} \frac{dt}{t} e^{-\pi (\rho + \chi) t}
\nonumber \\
&=& \Gamma \left[ 0, \pi \xi (\chi + \rho) \right]
\eeqn
where $\rho = \frac{\rho_1^2}{\Rf^2 \mu^2} + \frac{\rho_2^2}{\Rs^2 \mu^2}$
\item n=0, m $\ne$ 0
\beqn
\sum_{m \in Z}^{'} \int_{\xi}^{\infty} \frac{dt}{t} e^{- \pi t \frac{\frac{(m +
\rho_1)^2}{\Rf^2} + \frac{\rho_2^2}{\Rs^2}}{\mu^2}} e^{-\pi \chi t} &=& \sum_{m
\in Z}^{'} \int_{\xi \nu_1}^{\infty} \frac{dt}{t} e^{- \pi t (m + \rho_1)^2}
e^{-\pi \frac{\delta_1}{\nu_1} t} \nonumber \\
&=& {\cal R}_1 \left[\xi \nu_1, \rho_1, \frac{\delta_1}{\nu_1} \right]
\eeqn
where, $\nu_{1} = \frac{1}{\mu^2 \Rf^2}$ and $\delta_1 = \chi +
\frac{\rho_2^2}{\mu^2 \Rs^2} $.
\item m=0, n $\ne$ 0\\
Similar to the previous case with some parameters interchanged, we have:
\beqn
\sum_{n \in Z}^{'} \int_{\xi}^{\infty} \frac{dt}{t} e^{- \pi t
\frac{\frac{\rho_1^2}{\Rf^2} + \frac{(n+\rho_2)^2}{\Rs^2}}{\mu^2}} e^{-\pi \chi
t} &=& \sum_{n \in Z}^{'} \int_{\xi \nu_2}^{\infty} \frac{dt}{t} e^{- \pi t (n +
\rho_2)^2} e^{-\pi \frac{\delta_2}{\nu_2} t} \nonumber \\
&=& {\cal R}_1 \left[\xi \nu_2, \rho_2, \frac{\delta_2}{\nu_2} \right]
\eeqn
where, $\nu_{2} = \frac{1}{\mu^2 \Rs^2}$ and $\delta_2 = \chi +
\frac{\rho_1^2}{\mu^2 \Rf^2} $
\item Since the spectrum we are interested in has states that live either at odd
or even integers, it is write down the result of this integral in these limit
that the summation is over either even or odd integers:\\
\textit{n=0, m $\ne$ 0; m=even}\\
\beqn
{\cal R}_1^E \left[\xi \nu_1, \rho_1, \frac{\delta_1}{\nu_1} \right] &=& \sum_{m
\in Z}^{'} \int_{\xi}^{\infty} \frac{dt}{t} e^{-\pi t \frac{\frac{(2m +
\rho_1)^2}{\Rf^2} + \frac{\rho_2^2}{\Rs^2}}{\mu^2}} e^{-\pi \chi t}\nonumber \\
&=& \sum_{m \in Z}^{'} \int_{\xi}^{\infty} \frac{dt}{t} e^{-4 \pi t
\frac{\frac{(m + \frac{\rho_1}{2})^2}{\Rf^2} + \frac{\rho_2^2}{4\Rs^2}}{\mu^2}}
e^{-\pi \chi t/4}\nonumber \\
&=&\sum_{m \in Z}^{'} \int_{4 \xi \nu_1}^{\infty} \frac{dt}{t} e^{- \pi t (m +
\rho_1)^2} e^{-\pi \frac{\delta_1}{4 \nu_1} t} \nonumber \\
&=& {\cal R}_1 \left[4\xi \nu_1, \frac{\rho_1}{2}, \frac{\delta_1}{4 \nu_1}
\right]
\eeqn
where, $\nu_{1} = \frac{1}{\mu^2 \Rf^2}$ and $\delta_1 = \chi +
\frac{\rho_2^2}{\mu^2 \Rs^2} $ is the same as previously defined.\\[8pt]
\textit{n=0, m $\ne$ 0; m=odd}\\
\beqn
 {\cal R}_1^O \left[\xi \nu_1, \rho_1, \frac{\delta_1}{\nu_1} \right] &=&\sum_{m
\in Z} \int_{\xi}^{\infty} \frac{dt}{t} e^{- \pi t
\frac{\frac{(2m-1+\rho_1)^2}{\Rf^2} + \frac{\rho_2^2}{\Rs^2}} {\mu^2}} e^{-\pi
\chi t} \nonumber \\
&=&\sum_{m \in Z} \int_{\xi}^{\infty} \frac{dt}{t} e^{- 4 \pi t
\frac{\frac{(m+\frac{\rho_1-1}{2})^2}{\Rf^2} + \frac{\rho_2^2}{4\Rs^2}} {\mu^2}}
e^{-\pi \chi t/4} \nonumber \\
&=&  \int_{\xi \nu_1}^{\infty} \frac{dt}{t} e^{- \pi t \frac{(\rho_1-1)^2}{4}}
e^{-\pi \frac{\delta_1}{4 \nu_1} t} \nonumber \\ && \quad + \sum_{m \in Z}^{'}
\int_{4 \xi \nu_1}^{\infty} \frac{dt}{t} e^{- \pi t (m + \frac{\rho_1-1}{2})^2}
e^{-\pi \frac{\delta_1}{4 \nu_1} t} \nonumber \\ &=& \Gamma \left[ 0, \pi \xi
(\nu_1 (\rho_1 - 1)^2 + \delta_1) \right]+ {\cal R}_1 \left[4\xi \nu_1,
\frac{\rho_1-1}{2}, \frac{\delta_1}{4 \nu_1} \right]  \nonumber \\
\eeqn
In order to write the result of the integral in terms of the original ${\cal
R}_1$, we separate the zeroth term from the rest in the summation. It is also
useful to note that the function ${\cal R}_1$ is even in $\rho$ and hence:
\beq
{\cal R}_1 \left[ \xi, \rho, \delta \right] = {\cal R}_1 \left[ \xi, -\rho,
\delta \right]
\eeq
\end{itemize}

\section{Useful Limits of Relevant Functions}
\label{function_limits}
The result of the Kaluza-Klein integrals were evaluated in the previous section,
in terms of the two functions, $\Gamma[0, \pi \xi \chi]$ and ${\cal R}_1
\left[\xi, \rho, \delta \right]$. $\chi$ and $\xi$ are the regulators and in the
limit that they are zero, we can replace them with the relevant mass scales.
\begin{align}
 Q^2 \equiv \pi e^\gamma \chi \mu^2 \Big|_{\chi \rightarrow 0} \qquad \Lambda^2
\equiv \frac{\mu^2}{\xi}\Big|_{\xi \rightarrow 0}
\end{align}
As evaluated in the previous section:
\beqn
{\cal R}_1 \left[\xi, \rho, \delta \right] &=& - \Gamma \left[0, \pi \xi (\delta
+ \rho^2) \right] + \frac{2 e^{-\pi \delta \xi}}{\sqrt{\xi}} + 2 \pi
\sqrt{\delta} \text{Erf} \left[\sqrt{\pi \delta \xi}\right] \nonumber \\ &&
\qquad - \text{log} |2\ \text{sin} \pi (\rho + i \sqrt{\delta})|^{2} \nonumber
\eeqn
We use the following expansions:
\beqn
- \Gamma\left[0, z\right] &=& \gamma + \text{ln}\ z + \sum_{k \ge 1}
\frac{(-z)^k}{k!\ k} \qquad z > 0\\
\text{Erf} \left[x \right] &=& \frac{2x}{\sqrt{\pi}} - \frac{2x^3}{3\sqrt{\pi}}
+ {\cal O} (x^5) \qquad x \ll 1
\eeqn
Then,
\beqn
\Gamma \left[0, \pi \xi \chi \right] &=& -\gamma - \text{ln}\ \pi \xi \chi
\nonumber \\
&=&   - \text{ln}\ \pi \frac{e ^\gamma \mu^2 Q^2}{\Lambda^2 \pi e^\gamma \mu^2}
\nonumber \\
&=&  - \text{ln}\ \frac{Q^2}{\Lambda^2}
\eeqn
With these approximations, ${\cal R}_1 \left[\xi, \rho, \delta \right]$
simplifies to:
\beqn
{\cal R}_1 \left[\xi, \rho, \delta \right] &=& - \text{ln} \left[ 4 \pi
e^{-\gamma} \frac{1}{\xi} e^{-2/\sqrt{\xi}} \right] - \text{ln}
\left|\frac{\text{sin} (\rho + i \sqrt{\delta})}{\pi (\rho + i \sqrt{\delta})}
\right|^2
\eeqn
We summarize, the various terms that come up in the calculation of threshold
corrections in Section \ref{unification}. In the expressions below, we have also
introduced the compactifications scales $M_5 = \sqrt{\pi e^\gamma}/\Rf$ and $M_6
= \sqrt{\pi e^\gamma}/\Rs$.
\beqn
 \Gamma \left[0, \pi \xi \chi \right] &=& \text{ln} \frac{\Lambda^2}{Q^2}
\nonumber \\
 \Gamma \left[0, \pi \xi \nu_1 \right] &=& -\gamma - \text{ln}
\frac{\pi}{\Lambda^2 \Rf^2}\nonumber \\
&=& - \text{ln} \left[\frac{M_5^2}{\Lambda^2} \right]\nonumber \\
\Gamma \left[0, \pi \xi \nu_2 \right] &=& -\gamma - \text{ln}
\frac{\pi}{\Lambda^2 \Rs^2}\nonumber \\
&=& - \text{ln} \left[\frac{M_6^2}{\Lambda^2} \right]\nonumber \\
{\cal R}_1 \left[4 \xi \nu_1, 0,  \frac{\chi}{4 \nu_1} \right] &=& -\text{ln}
\left[\pi e^{-\gamma - \Lambda \Rf} (\Lambda \Rf)^2\right] \nonumber \\
{\cal R}_1 \left[4 \xi \nu_1, \frac{1}{2},  \frac{\chi}{4 \nu_1} \right] &=&
-\text{ln} \left[\pi e^{-\gamma - \Lambda \Rf} (\Lambda \Rf)^2\right] -
\text{ln} \left[\frac{2}{\pi} \right]^2 \nonumber \\
{\cal R}_1 \left[4 \xi \nu_2, 0,  \frac{\chi}{4 \nu_2} \right] &=& -\text{ln}
\left[\pi e^{-\gamma - \Lambda \Rs} (\Lambda \Rs)^2\right] \nonumber \\
{\cal R}_1 \left[4 \xi \nu_2, \frac{1}{2},  \frac{\chi}{4 \nu_2} \right] &=&
-\text{ln} \left[\pi e^{-\gamma - \Lambda \Rs} (\Lambda \Rs)^2\right] -
\text{ln} \left[\frac{2}{\pi} \right]^2 \nonumber \\
\Gamma \left[0, \pi \xi \left(\frac{\nu_1}{4} + \frac{\nu_2}{4} \right) \right]
&=&  -\text{ln} \left[\frac{M_5^2 + M_6^2}{4 \Lambda^2} \right]\nonumber \\
\eeqn

\section{6D $\rightarrow$ 4D matching}
\label{analyticalsolution}
We calculated the corrections to the gauge couplings coming from the KK states
of the 6D orbifold model
that was constructed. At the lowest compactification scale (largest
compactification radius), we said that
the couplings from 4D MSSM and 6D orbifold model should match. In this section,
we will compare the two
sets of equations, from the two theories:
\beqn
 \alpha_{i}^{-1} (Q) &=& \alpha^{-1} (\Lambda) +
\sum_{\rho} \Omega_{i, \rho} (Q) \nonumber \\
 \alpha_{i}^{-1} (Q) &=& \alpha_{GUT}^{-1} + \frac{b_{i}}{2\pi} log
\frac{M_{GUT}}{Q} - \alpha_{GUT}^{-1} \frac{\epsilon_3}{(1 + \epsilon_3)}
\delta_{i3} \nonumber
\eeqn
and solve for the three scales of the orbifold model, $\Lambda$, $M_5$, and
$M_6$
as well as coupling constant, $\alpha$ at the cut-off scale.

Since the two expressions have to match at all scales below the smallest
compactification scale of the orbifold model,
we can rewrite the above two equations as:
\beqn
\alpha_{GUT}^{-1} &+& \frac{b_{i}}{4\pi} \text{ln}
\frac{M_{GUT}^2}{Q^2} - \alpha_{GUT}^{-1} \frac{\epsilon_3}{(1 + \epsilon_3)}
\delta_{i3}  \nonumber \\
&=&  \alpha^{-1} (\Lambda) +  \frac{b_i^{++} (\irho = 0)}{4 \pi} \text{ln}
\frac{\Lambda^2}{Q^2} + \left(\frac{b_i^{+-} (\irho = 0) + b_i^{-+} (\irho =
0)}{4 \pi}\right) \text{ln} \left[\frac{\pi \Lambda}{2 M_5}
\right]^2  \nonumber \\
&+& \left(\frac{b_i^{+-} (\irho = 0) + b_i^{--} (\irho = 0)}{4 \pi}\right)
\text{ln} \left[\frac{\pi \Lambda}{2 M_6} \right]^2 + \frac{b_i^{+-} (\irho =
2)} {4 \pi} \text{ln} \left[ \frac{4 \Lambda^2}{ M_5^2+
M_6^2}\right] \nonumber \\
\eeqn
where we have used the complete expression we estimated for the corrections to
couplings in \ref{corrections}.

We use the following redefinitions:
\begin{align}
 \frac{b_i^{MSSM}}{4 \pi} =  \frac{b_i^{++}(\irho = 0)}{4 \pi} &=& \beta_i
\nonumber \\
\frac{b_i^{+-} (\irho = 0) + b_i^{-+} (\irho =0)}{4 \pi} &=& -A_{i} \nonumber \\
\frac{b_i^{+-} (\irho = 0) + b_i^{--} (\irho = 0)}{4 \pi} &=& -B_i \nonumber \\
\frac{b_i^{+-} (\irho = 2)} {4 \pi} &=& -C_i
\end{align}
and
\begin{align}
(A_i + B_i) \text{ln} \left[\frac{\pi}{2} \right]^2 +C_i \text{ln} \left[ 4
\right] = D_i
\end{align}
and hence end up with a set of three equations that can be simply written as:
\beqn
\alpha_{GUT}^{-1} &-& \alpha^{-1} (\Lambda)- \alpha_{GUT}^{-1}
\frac{\epsilon_3}{(1 + \epsilon_3)}
\delta_{i3} -\beta_i \text{ln}
\frac{\Lambda^2}{M_{GUT}^2}   \nonumber \\
&+&  A_i \text{ln}
\frac{\Lambda^2}{M_5^2} + B_i\text{ln} \frac{\Lambda^2}{M_6^2} + C_i\text{ln}
\frac{\Lambda^2}{ M_5^2+
M_6^2} + D_i = 0
\label{simple}
\eeqn
where, $A_i = A_{1i} + A_{2i}$ and $i = 1,2,3$. We look at the equations
corresponding (i) (i = 1) - (i =2)
(ii) i = 2 (iii) i = 3 and solve for $\Lambda$, $M_5$, and $M_6$. It is usually
considered that the 4D unification
scale is around $3.0 \times 10^{16}$ GeV and the couplings at this scale are
unified at $\alpha_{GUT}^{-1} = 24$.
In standard scenarios of MSSM with gaugino mass unification, $\epsilon_3 = -3
\%$. Depending of the spectrum of low energy
SUSY, these quantities are subject to change. The first equation we get by
simplifying Eq. (\ref{simple}) for (i = 1) - (i =2)
is:
\beqn
-(\beta_1 - \beta_2) \text{ln}\frac{\Lambda^2}{M_{GUT}^2} &+& (A_1-A_2)
\text{ln}
\frac{\Lambda^2}{M_5^2} + (B_1-B_2)\text{ln} \frac{\Lambda^2}{M_6^2} \nonumber
\\ &+& (C_1-C_2)\text{ln} \frac{\Lambda^2}{ M_5^2+
M_6^2} + (D_1-D_2) = 0
\eeqn
Defining, $\frac{\Lambda^2}{M_5^2} = X$ and $\frac{\Lambda^2}{M_6^2} = Y$, we
get:
\beqn
\text{ln}\frac{\Lambda^2}{M_{GUT}^2} &=& \left(\frac{A_1-A_2}{\beta_1 -
\beta_2}\right) \text{ln} X +
 \left(\frac{B_1-B_2}{\beta_1 - \beta_2}\right)\text{ln} Y \nonumber \\
 &-& \left(\frac{C_1-C_2}{\beta_1 - \beta_2}\right)\text{ln} \left( \frac{1}{X}
+ \frac{1}{Y} \right)
+\left(\frac{D_1-D_2}{\beta_1 - \beta_2}\right)
\label{mgeq}
\eeqn
Next, we look at Eq (\ref{simple}) when $i=2$:
\beqn
\alpha_{GUT}^{-1} &-& \alpha^{-1} (\Lambda)-\beta_2
\text{ln}\frac{\Lambda^2}{M_{GUT}^2}   \nonumber \\
&+&  A_2 \text{ln} X + B_2\text{ln} Y - C_2\text{ln} \left(\frac{1}{X} +
\frac{1}{Y} \right) + D_2 = 0
\eeqn
Then, using the expression we just derived in Eq. (\ref{mgeq}), we get an
expression for $\alpha^{-1} (\Lambda)$:
\beqn
 \alpha^{-1} (\Lambda) &=& \alpha_{GUT}^{-1} + \left[A_2 - \beta_2
\left(\frac{A_1-A_2}{\beta_1-\beta_2}\right) \right]
 \text{ln} X + \left[B_2 - \beta_2 \left(\frac{B_1-B_2}{\beta_1-\beta_2}\right)
\right]\text{ln} Y \nonumber \\
&-& \left[C_2 - \beta_2 \left(\frac{C_1-C_2}{\beta_1-\beta_2}\right)
\right]\text{ln} \left(\frac{1}{X} + \frac{1}{Y} \right)
+ \left[D_2 - \beta_2 \left(\frac{D_1-D_2}{\beta_1-\beta_2}\right) \right]
\nonumber \\
\label{invastr}
\eeqn
Finally, we look at Eq (\ref{simple}) when $i=3$, and simplify it using the
relations obtained in Eqs.
(\ref{mgeq} \& \ref{invastr}) and we get a final expression:
\beqn
\left[ A_3-A_2 +
\left(\frac{A_1-A_2}{\beta_1-\beta_2}\right)(\beta_2-\beta_3)\right] \text{ln} X
&+& \left[ B_3-B_2 +
\left(\frac{B_1-B_2}{\beta_1-\beta_2}\right)(\beta_2-\beta_3)\right] \text{ln} Y
\nonumber \\ - \left[ C_3-C_2 +
\left(\frac{C_1-C_2}{\beta_1-\beta_2}\right)(\beta_2-\beta_3)\right] \text{ln}
\left(\frac{1}{X} + \frac{1}{Y} \right) &+& \left[ D_3-D_2 +
\left(\frac{D_1-D_2}{\beta_1-\beta_2}\right)(\beta_2-\beta_3)\right] \nonumber
\\ - \alpha^{-1}_{GUT} \frac{\epsilon_3}{1+\epsilon_3} &=& 0
\label{final}
\eeqn
The above three equations can be rewritten in a simple manner as (in the order
Eq. (\ref{final}), (\ref{mgeq}), (\ref{invastr})):
\beqn
{\cal A} \text{ln} X + {\cal B} \text{ln} Y - {\cal C} \text{ln}
\left(\frac{1}{X} + \frac{1}{Y} \right) + {\cal D} &=& 0 \nonumber \\
{\cal F} \text{ln} X + {\cal G} \text{ln} Y - {\cal H} \text{ln}
\left(\frac{1}{X} + \frac{1}{Y} \right) + {\cal I} &=& \text{ln}
\frac{\Lambda^2}{M_{GUT}^2} \nonumber \\
{\cal K} \text{ln} X + {\cal L} \text{ln} Y - {\cal M} \text{ln}
\left(\frac{1}{X} + \frac{1}{Y} \right) + {\cal N} &=& \alpha^{-1} (\Lambda)
\nonumber \\
\label{threeeq}
\eeqn
with,
\begin{small}
\begin{align}
{\cal A} &=& A_3 - A_2 + (A_1 - A_2) \left(\frac{\beta_2 - \beta_3}{\beta_1 -
\beta_2} \right), \nonumber \\
{\cal B} &=& B_3 - B_2 + (B_1 - B_2) \left(\frac{\beta_2 - \beta_3}{\beta_1 -
\beta_2} \right),  \nonumber \\
{\cal C} &=& C_3 - C_2 + (C_1 - C_2) \left(\frac{\beta_2 - \beta_3}{\beta_1 -
\beta_2} \right), \nonumber \\
{\cal D} &=& D_3 - D_2 + (D_1 - D_2) \left(\frac{\beta_2 - \beta_3}{\beta_1 -
\beta_2} \right) - \alpha_{GUT}^{-1} \frac{\epsilon_3}{(1 + \epsilon_3)}
\nonumber \\
{\cal F} &=& \frac{A_1 - A_2}{\beta_1 - \beta_2}, \nonumber \\
{\cal G} &=& \frac{B_1 - B_2}{\beta_1 - \beta_2}, \nonumber \\
{\cal H} &=& \frac{C_1 - C_2}{\beta_1 - \beta_2}, \nonumber \\
{\cal I} &=& \frac{D_1 - D_2}{\beta_1 - \beta_2}, \nonumber \\
{\cal K} &=& A_2- \beta_2 \left(\frac{A_1 - A_2}{\beta_1 - \beta_2} \right),
\nonumber \\
{\cal L} &=& B_2- \beta_2 \left(\frac{B_1 - B_2}{\beta_1 - \beta_2} \right),
\nonumber \\
{\cal M} &=& C_2- \beta_2 \left(\frac{C_1 - C_2}{\beta_1 - \beta_2} \right),
\nonumber \\
{\cal N} &=& D_2- \beta_2 \left(\frac{D_1 - D_2}{\beta_1 - \beta_2} \right) +
\alpha^{-1}_{GUT}, \nonumber
\end{align}
\end{small}
These quantities can be calculated using the beta-function coefficients given in
Table. \ref{finalbeta}. The numerical values of all the above coefficients are
summarized in Table \ref{cal}.

\begin{table}
\begin{center}
 \begin{tabular}{|c|c|c|}
\hline
  Coefficient & Value\\
\hline
${\cal A}$ & $-\frac{3}{7\pi}$\\
${\cal B}$ & $\frac{6}{7 \pi}$ \\
${\cal C}$ & $-\frac{3}{7 \pi}$\\
${\cal D}(\epsilon_3)$ & $-\iag \frac{\epsilon_3}{1 + \epsilon_3} - \frac{6}{7
\pi} \text{ln} \frac{4}{\pi}$ \\
\hline
${\cal F}$ & $\frac{4}{7}$ \\
${\cal G}$ & $\frac{6}{7}$ \\
${\cal H}$ & $-\frac{3}{7}$ \\
${\cal I}$ & $-\frac{3}{7} \text{ln} 4 - \frac{20}{7} \text{ln} \frac{2}{\pi}$
\\
\hline
${\cal K}$ & $-\frac{9}{14\pi}$ \\
${\cal L}$ & $-\frac{12}{7\pi}$ \\
${\cal M}$ & $-\frac{9}{14 \pi}$ \\
${\cal N}$ & $\alpha^{-1}_{GUT}-\frac{6}{7\pi} \text{ln} 2 - \frac{33}{7\pi}
\text{ln} \frac{2}{\pi}$ \\
\hline
 \end{tabular}
\caption{\label{cal}The coefficients in the expression Eq. (\ref{threeeq})}
\end{center}
\end{table}
With these coefficients, we get a simple quadratic equation in terms in of the
variables X and Y:
\begin{equation}
 \left(\frac{Y}{X}\right)^2 + \frac{Y}{X} - \text{Exp}\ \left( -\frac{7 \pi
{\cal D}(\epsilon_3)}{3}\right) =0
\end{equation}
Recall that $X = \frac{\Lambda^2}{M_5^2}$ and $Y = \frac{\Lambda^2}{M_6^2}$,
which implies that the above equation turns into a quadratic equation in
$\left(\frac{M_5}{M_6}\right)^2$ with the solution.
\begin{align}
M_5^2 &=& \frac{-1 \pm \sqrt{1+4 \text{Exp}\ \left( -\frac{7 \pi {\cal
D}(\epsilon_3)}{3}\right)}} {2} M_6^2
\end{align}
which we write as $M_5 = \sqrt{m(\epsilon_3)}M_6$. The slope $m$, is the
positive solution from the above expression and is shown in Fig. \ref{me3}. The
other two equations then yield us $M_5$ and $M_6$ uniquely and one expression
relating $\alpha^{-1}(\Lambda)$ and $\Lambda$.
\beqn
M_5 &=& \left( m(\epsilon_3)^{({\cal G- H})/2} (m(\epsilon_3)+1)^{{\cal H}/2}
e^{{\cal I}/2}\right) \mgut \nonumber \\
M_6 &=& \left( m(\epsilon_3)^{({\cal G- H}-1)/2} (m(\epsilon_3)+1)^{{\cal H}/2}
e^{{\cal I}/2}\right) \mgut \nonumber \\
\alpha^{-1}(\Lambda) &=& - \frac{3}{\pi} \text{ln} \frac{\Lambda^2}{M_{GUT}^2} +
\frac{3}{\pi} \text{ln} \left(m(\epsilon_3)^{({\cal G- H})}
(m(\epsilon_3)+1)^{{\cal H}}  e^{{\cal I}} \right) \nonumber \\ &+& \text{ln}
\left(m(\epsilon_3)^{({\cal L- M})} (m(\epsilon_3)+1)^{{\cal M}}  e^{{\cal N}}
\right)
\eeqn

These expressions are plotted in figure.

\end{document}